\documentclass[preprint,showpacs,amsmath,amssymb,aip,apl]{revtex4-1} 

\usepackage{graphicx}
\usepackage{dcolumn}
\usepackage{bm}
\usepackage{braket}
\usepackage{mathtools}
\usepackage[utf8]{inputenc}
\usepackage[T1]{fontenc}
\usepackage{mathptmx}
\usepackage{hyperref}
\usepackage{color}

\begin{document}

\title{Demonstration of NV-detected ESR spectroscopy at 115 GHz and 4.2 Tesla}

\author{Benjamin Fortman}
\affiliation{Department of Chemistry, University of Southern California, Los Angeles CA 90089, USA}

\author{Junior Pena}
\affiliation{Department of Physics \& Astronomy, University of Southern California, Los Angeles CA 90089, USA}

\author{Karoly Holczer}
\affiliation{Department of Physics \& Astronomy, University of California, Los Angeles CA 90095, USA}

\author{Susumu Takahashi}
\email{susumu.takahashi@usc.edu}
\affiliation{Department of Chemistry, University of Southern California, Los Angeles CA 90089, USA}
\affiliation{Department of Physics \& Astronomy, University of Southern California, Los Angeles CA 90089, USA}

\date{\today}

\begin{abstract}
High frequency electron spin resonance (ESR) spectroscopy is an invaluable tool for identification and characterization of spin systems. Nanoscale ESR using the nitrogen-vacancy (NV) center has been demonstrated down to the level of a single spin. However, NV-detected ESR has exclusively been studied at low magnetic fields, where spectral overlap prevents clear identification of spectral features. 
Within this work, we demonstrate NV-detected ESR measurements of single-substitutional nitrogen impurities in diamond at a NV Larmor frequency of 115 GHz and the corresponding magnetic field of 4.2 Tesla. 
The NV-ESR measurements utilize a double electron-electron resonance sequence and are performed using both ensemble and single NV spin systems.
In the single NV experiment, chirp pulses are used to improve the population transfer and for NV-ESR measurements. 
This work provides the basis for NV-based ESR measurements of external spins at high magnetic fields.

\end{abstract}

\maketitle

%
%
The nitrogen-vacancy (NV) center has unique properties that make it an excellent candidate for high sensitivity magnetic sensing.~\cite{Degen2008, Maze2008, Balasubramanian2008}
The NV center is a two atom defect in the diamond lattice, with the capacity for optical spin-state initialization and readout, long coherence times, and high sensitivity to external magnetic fields.\cite{Gruber1997, balasubramanian08,Wolf2015}
NV-detected electron spin resonance (ESR) offers the capability to detect a single or a small number of electron spins~\cite{Grinolds2013, DeLange2012, Abeywardana2016, Shi2013, Schlipf2017, Sushkov2014} and to investigate biological molecules at the single molecule level.~\cite{Shi2015, Shi2018} 
Such an ESR technique with single spin sensitivity potentially eliminates ensemble averaging in heterogeneous and complex systems, and has great promise to directly probe fundamental interactions and biochemical function.
In ESR, the measurement of the g-factor is extremely useful for the identification of spin species. 
However, a featureless "g = 2" signal is often observed, causing spectral overlap with target ESR signals, which may prevent spin identification.~\cite{Mamin2012, Grotz2011, Shi2018, meriles12}

Similar to nuclear magnetic resonance (NMR) spectroscopy, pulsed ESR spectroscopy at higher frequencies (HF) and magnetic fields becomes more powerful for finer spectral resolution, enabling clear spectral separation of systems with similar g values.~\cite{Savitsky2009, Krzystek2006} 
This is advantageous in the investigation of complex and heterogeneous spin systems.~\cite{Hubbell2000, Lawrence1999}
A high frequency of Larmor precession is also less sensitive to motional narrowing, enabling the ESR investigation of structures for molecules in motion.~\cite{freed11, schweiger2001principles}
In addition, a high Larmor frequency provides greater spin polarization: improving sensitivity~\cite{Savitsky2009, Krzystek2006} and providing control of spin dynamics.~\cite{Takahashi08, Takahashi11}
On the other hand, pulsed HF ESR often has the disadvantage of long pulse times due to low HF microwave power. 
The low microwave power limits the excitation bandwidth, and consequently the sensitivity of pulsed ESR measurements.
NV-detected ESR (indicated as NV-ESR) will overcome this limitation and improve the sensitivity of HF ESR drastically.
However, only a few investigations of NV centers have been performed at high magnetic fields,~\cite{Stepanov2015, Aslam2015, Aslam2017} and NV-ESR has not been demonstrated at a high magnetic field. 

In this work, we demonstrate NV-ESR at a Larmor frequency of 115 GHz, corresponding to a magnetic field of $\sim$4.2 Tesla.
The HF NV-ESR experiment is performed with both an ensemble and a single NV system.
Within the ensemble experiment, we start the characterization of NV centers using optically detected magnetic resonance (ODMR), a measurement of Rabi oscillations, and a spin echo measurement to determine a spin decoherence time (\textit{T}$_2$).
Then, we utilize a double electron-electron resonance (DEER) sequence to perform NV-ESR spectroscopy of single-substitutional nitrogen (P1) centers in diamond. 
We find that the observed NV-ESR spectrum is in excellent agreement with the spectrum of P1 centers.
In the single NV-ESR experiment, we start the identification and the characterization of a single NV center.
For high fidelity coherent control, we apply chirp pulses which improve population inversion and optical contrast.
We then implement a DEER sequence with chirp pulses.
The observed NV-ESR signal is in agreement with P1 centers. 
This work provides a clear demonstration of HF NV-detected ESR, and provides a foundation for the study of external spins with high spectral resolution NV-ESR. 
Furthermore, the presented experimental strategies are applicable to NV-ESR at higher magnetic fields.

%
%
\begin{figure}
    \centering
    \includegraphics[width=3.3in]{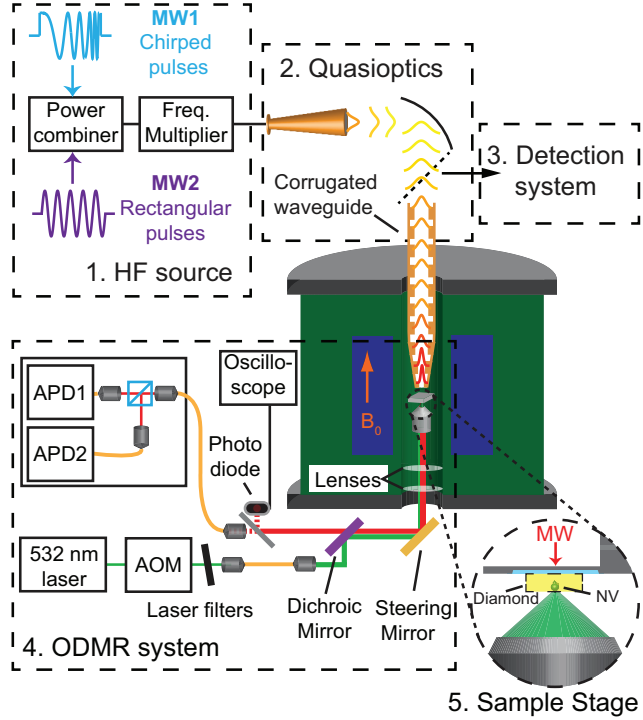}
    \caption{Overview of HF ODMR system. The HF ODMR system consists of five components. (1) A HF microwave component, (2) a quasioptical MW propagation system, (3) a HF ESR detection system, (4) an ODMR system. The NV detection system utilizes a photodiode for ensemble experiments and avalanche photodiodes for single NV experiments.
    (5) A diamond sample is mounted upon a sample stage within a variable field magnet. 
    HF MW excitation propagates through free space from the waveguide to the top of the sample stage. Optical access is obtained through the bottom of the sample stage.
}
    \label{fig:Setup}
\end{figure}
Figure~\ref{fig:Setup} shows an overview of a home-built HF ODMR system used in the experiment. 
The HF ODMR system consists of a HF microwave source (Virginia Diodes, Inc.), quasioptics, a 12.1 Tesla cryogenic-free superconducting magnet (Cryogenics), and a confocal microscope system for ODMR. The HF ODMR system is built upon the existing HF ESR spectrometer that was described previously.\cite{Cho2014, Cho2015} Therefore the system enables \textit{in-situ} experiments of both ESR and ODMR. 
As seen in Fig. 1, the HF source contains two microwave synthesizers (MW1 and MW2), a power combiner, and a frequency multiplier chain. An IQ mixer (Miteq) controlled by an arbitrary wave generator (AWG; Keysight) has recently been implemented in MW1 for pulse shaping of high-frequency microwaves. Two synthesizers are employed for DEER experiments.
The frequency range of the microwave source is 107-120 GHz and 215-240 GHz. 
In this experiment, we use a frequency range of 107-120 GHz where the output power of the HF microwave is 480 mW at 115 GHz. 
HF microwaves are propagated to a sample using a home-built quasioptical bridge and a corrugated waveguide (Thomas-Keating). 
As demonstrated previously, quasioptics are suitable for a high-frequency ESR spectrometer because of their capacity for low-loss and broadband propagation.~\cite{Smith98,VanTol2005}
A sample is mounted at the end of the corrugated waveguide and positioned at the field center of a room temperature bore within a superconducting magnet system. 
No microwave resonator is employed for implementation of wide bandwidth DEER techniques.
The magnetic field at the sample is adjustable between 0 to 12.1 Tesla. For the single-NV HF ODMR experiment, we employ a conventional confocal microscope setup routinely used for NV ODMR experiments.~\cite{Abeywardana2017, Fortman2019} The details of the single-NV ODMR system have been described previously.~\cite{Stepanov2015}
For the ensemble HF ODMR experiment, we direct the fluorescence (FL) to a photodiode (Thorlabs), implemented before the coupling stage to the single mode fiber. This is connected to a fast oscilloscope (Le Croy) for measurement of the time-domain FL signal.
The detection volume in the ensemble experiment is in the range of a few $\mu$m$^3$.
Results presented here have been obtained using two samples, both are 2.0 x 2.0 x 0.3 mm$^3$ size, (111)-cut high-pressure, high-temperature type-Ib diamonds from Sumitomo Electric Industries. The crystal used for the single NV experiment had been previously shown to contain single NV centers with reasonably long spin-relaxation times and coupling to P1 centers.~\cite{Fortman2019}
The other crystal, used for the ensemble experiment, was subjected to successive irradiations with high energy (4 MeV) electron beam and annealing processes (at 1000 $^o$C) in order to increase the NV center density. 
Exposure to a total fluence of $1.2\times10^{18}$ e$^{-}$/cm$^2$ resulted in approximately 8{\%} NV/N ratio as determined from the X-band ($\sim$9 GHz) ESR spectrum of the sample (see the supplementary material). 
Characterization with ensemble HF ESR measurements reveals strong ESR signals from both P1 and NV centers, indicating that both NV and P1 concentrations are more than 1 ppm.

\begin{figure}
    \centering
    \includegraphics[height=4in]{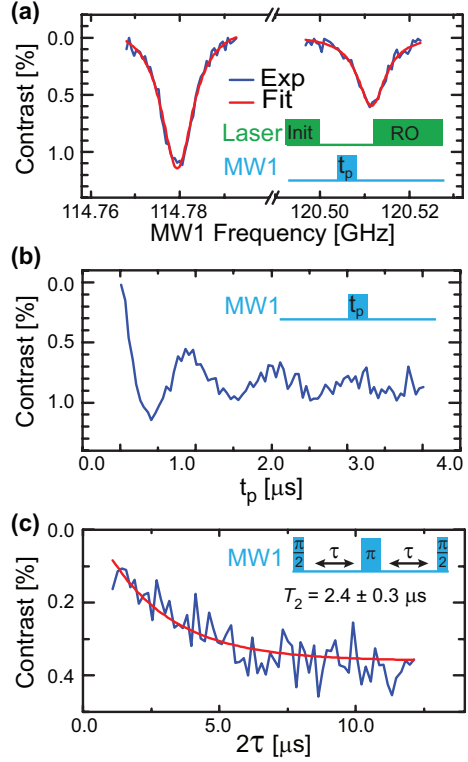}
    \caption{Characterization of an ensemble NV system at 115 GHz and 4.2 Tesla.
    (a) Pulsed ODMR signals of NV centers. A pulse length ($t_p$) of 500 ns was used. 
    To confirm the observed signal, sequential measurements were made with (Sig) and without (Ref) application of MW irradiation. The contrast in percentage (Exp = (Ref-Sig)/Ref) was plotted for the analysis.
    A Lorentzian fit is shown in red.
    Laser pulses with durations of $5$ and $10$ $\mu$s were used for initialization (Init) and readout (RO), respectively.     
    (b) Rabi oscillations measurement. 
    The data was taken by varying the length of $t_p$. 
    From the data, $\pi/2$ and $\pi$ pulse lengths of $212$ and  $402$ ns  were obtained.
    (c) Spin echo decay data obtained by varying the inter-pulse delay time $\tau$. 
    $T_2$ was determined to be $2.4 \pm 0.3 $ $\mu$s when fit to a single exponential decay.
    The pulse sequences were shown in the inset.
    All ensemble experiments were repeated $\sim 10^3$ times for averaging.
    }
    \label{fig:NVData}
\end{figure}

The extracted pulse times were used in a spin echo experiment. A spin echo relaxation time of $2.4 \pm 0.3$ $\mu$s was observed, as seen in Fig. \ref{fig:NVData}(c).
\begin{figure}
    \centering
    \includegraphics[width=3.3in]{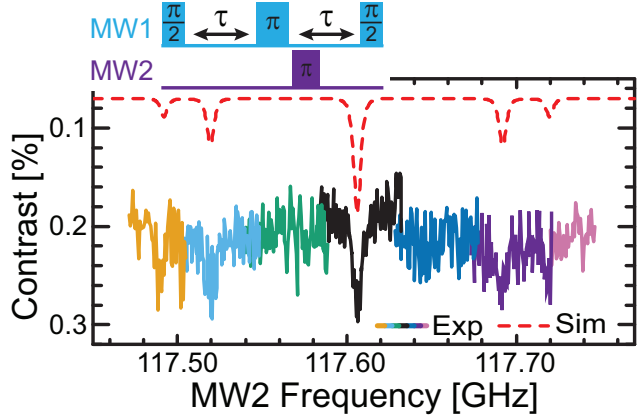}
    \caption{HF NV-ESR using an ensemble NV system. 
    A simulation of the spectrum based upon DEER spin dynamics is shown in the red dashed line.\cite{Stepanov2016}
    For the simulation, $B_0 = 4.1963$ Tesla with a polar angle of $1.8$ degrees from the from the NV center's [111] axis. Due to the small polar angle, sensitivity in azimuthal angle variation was below the observed linewidth. 
    The pulse sequence is shown in the inset.
    Each experimental scan is shown in a different color. 
    In this measurement, $\tau = 1$ $\mu$s and $\pi_{MW2} = 560$ ns were used.
    The simulation is offset for clarity.
    }
    \label{fig:HfEnsDeer}
\end{figure}
We next perform NV-ESR using the ensemble NV system with a DEER sequence. 
In the DEER sequence, a separate MW pulse (MW2) is 
applied during the spin echo sequence as shown in the inset of Fig.~\ref{fig:HfEnsDeer}. 
When the frequency of the pulse matches the ESR frequency of target spins, the target spins flip, and then the dipolar field from the target spins experienced by the NV center changes. 
This change results in a reduction of the refocused echo intensity.
In this manner, an ESR signal of weakly dipolar coupled spins located in the nanometer scale region surrounding the NV center can be detected.~\cite{DeLange2012,Abeywardana2016} 
As shown in  Fig.~\ref{fig:HfEnsDeer}, application of this pulse sequence reveals five distinct reductions in FL intensity at $117.49$, $117.52$, $117.61$, $117.69$, and $117.72$ GHz. These dips are in excellent agreement with the spectrum simulated from the P1 center's Hamiltonian ($S = 1/2$, $I =1$, $g = 2.0024$, $A_{\perp} = 82$ MHz, $A_{\parallel} = 114$ MHz).~\cite{Stepanov2016,Loubser1978}

\begin{figure*}
    \centering
    \includegraphics[height=3in]{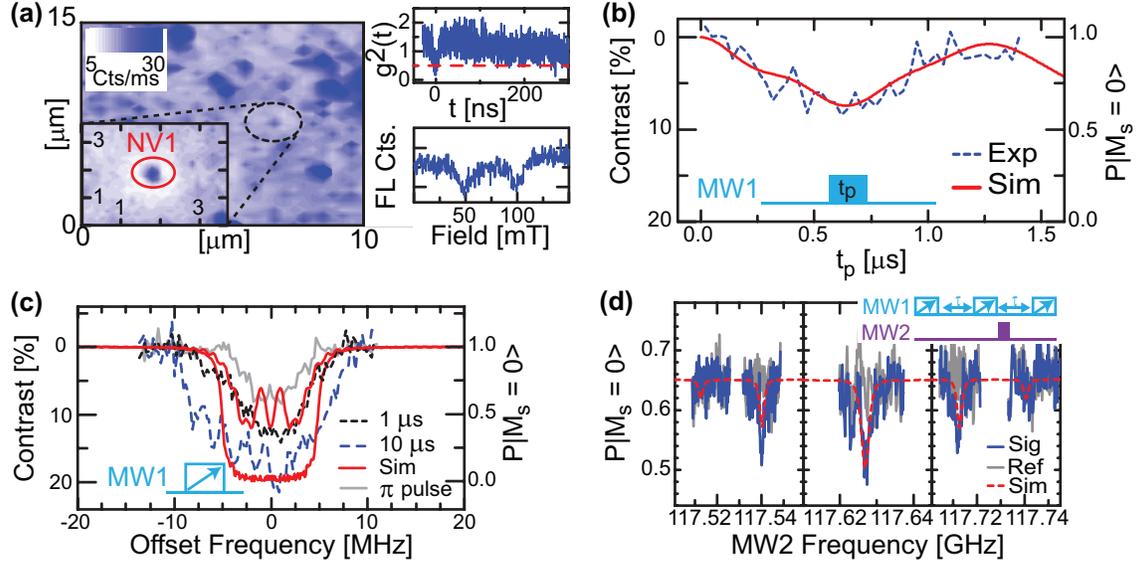}
    \caption{115 GHz NV-ESR spectroscopy of P1 centers using a single NV.
    (a) FL image of NV1. The top inset figure shows an autocorrelation measurement on NV1.
    The dashed red line in the top inset indicates the threshold for a single quantum emitter. 
    The bottom inset shows the FL from NV1 as a function of magnetic field. 
    Dips at $50$ and $100$ mT correspond to the level anti-crossing of the excited and ground states of the NV center, respectively.~\cite{Stepanov2015}
    (b) Rabi data taken at $4.197$ Tesla corresponding to the $m_S = 0 \leftrightarrow m_S=-1$ transition of NV1. 
    The normalized percentage of FL contrast (dashed blue line) is plotted against the pulse length.
    The simulation is shown in red. 
    For all single NV measurements, the signal was averaged for ($10^5 - 10^6$) scans with a 5 $\mu$s initialization and 300 ns readout pulse.
    (c) Pulsed ODMR data taken using a MW1 pulse. 
    These pulses were implemented experimentally by varying the position of the carrier frequency and sweeping from $1$ to $11$ MHz (after the multiplication chain).  
    The data is plotted with respect to the center resonance frequency. The simulations are shown in red for the $1$ $\mu$s and $10$ $\mu$s pulse lengths. 
    The effect of a monochromatic pulse is shown for comparison.
    (d) NV-ESR spectrum using NV1 taken with a MW1 chirp pulse length of $1 \mu$s, $\tau = 4$ $\mu$s, and $\pi_{MW2} = 1$ $\mu$s. The NV-ESR was measured by varying the frequency of MW2 (Sig). Clear reductions in FL intensity were observed at ESR frequencies of P1 centers compared to a reference with no MW2 pulse (Ref).
    The simulation is shown for P1 centers.}
    \label{fig:SingleNVData}
\end{figure*}

Having resolved the spectrum of P1 centers from the ensemble system, we next discuss the demonstration of NV-ESR using a single NV center. 
For this we begin by taking a FL image of the diamond, as is shown in  Fig. \ref{fig:SingleNVData}(a).
From the observed FL image, a well isolated FL spot is selected (denoted as NV1). An autocorrelation measurement performed on the FL spot, as shown in the upper inset of Fig.~\ref{fig:SingleNVData}(a), confirms the observed FL is from a single quantum emitter. 
As can be seen in the lower inset of Fig.~\ref{fig:SingleNVData}(a), upon increasing the magnetic field we observe the level anti-crossing of both the ground and excited state, thereby confirming the FL spot as a single NV center.~\cite{Epstein05, Stepanov2015} 
After the detection of a single NV center, the superconducting magnet was set in a persistent field mode at $\sim$ 4.2 Tesla. 
The magnetic field strength was then measured via an \textit{in-situ} ESR measurement of P1 centers.
Next we perform pulsed ODMR on the NV center and resolve clear reductions in FL intensity at both 114.80 and 120.53 GHz, corresponding to the lower and upper transitions (data not shown).
From these ODMR signals, we determined the NV center to be aligned in the magnetic field with a polar angle of 1.6 degrees from the [111] axis.
We next perform a measurement of the Rabi oscillations for NV1 at 114.80 GHz.
The FL intensity against pulse length is shown within Fig.~\ref{fig:SingleNVData}(b) and shows clear Rabi oscillations.
Analysis of the signal was then performed by simulating dynamics of a two-level system using the Liouville-von Neumann equation with a resonance-frequency distribution due to the hyperfine coupling of $^{14}$N ($2.2$ MHz).
As shown in Fig.~\ref{fig:NVData}(b), the experiment and the simulation show excellent agreement.
The analysis also reveals that the limited excitation bandwidth results in incomplete population inversion of the NV center spin state.
The estimated population inversion is only $\sim$ 40$\%$.

The small population inversion of a rectangular pulse is a significant challenge for the NV-ESR experiment because of errors in the preparation and readout of the quantum coherent state used in the DEER sequence.
The result can be poor signal-to-noise in the measurement.
To overcome this, we employ a pulse shaping technique.
The recent development of high temporal resolution (sub-ns) arbitrary waveform generators has triggered significant progress in various pulse shaping techniques for ESR. 
Here we focus on chirp pulses, a class of pulses that have been used to demonstrate broadband control over wide frequency ranges. 
Chirped pulses are now routinely used in ESR spectroscopy at X- and Q-(~34 GHz) bands.~\cite{Doll2013,Doll2014,Wili2018} 
In addition, it has recently been demonstrated at a frequency of 200 GHz.\cite{Kaminker2017} 
Chirp pulses offer wider spectral excitation, an ability to correct pulse imperfections, and generally higher fidelity than rectangular pulses.\cite{Tannus1997}
The effectiveness of chirp pulses is due to the principle of adiabatic passage, whereby population transfer is achieved by sweeping an applied electromagnetic field through resonance at a sufficiently slow rate.\cite{Abragam1961, Baum1985, Jeschke2015}

Here, we utilize a linear frequency-swept chirp pulse at 115 GHz.
Figure~\ref{fig:SingleNVData}(c) shows the efficiency of chirp pulses with two different frequency-swept rates; 10 MHz-swept in a duration of 1 $\mu$s and 10 $\mu$s. 
A clear increase in contrast is observed via the application of chirp pulses, indicating an improvement in both population transfer and excitation bandwidth compared with the rectangular pulse.
$T_2$ was also determined to be $\sim$ $20$ $\mu$s from a spin echo measurement (see the supplementary material).
Furthermore, we use the Liouville-von Neumann equation to calculate the observed behavior as a function of frequency based on the chosen pulse length and sweep width. As seen in Fig.~\ref{fig:SingleNVData}(c), the observed result is in good agreement with the simulation, confirming that the usage of chirp pulses increases population transfer for the NV center to nearly $100\%$. 

We next utilize chirp pulses to perform NV-ESR with the NV center by applying three pulses with a fixed delay between them.
Similarly to the previous work,~\cite{Fortman2019}
a long MW2 pulse ($\pi_{MW2} = 1$ $\mu$s) was utilized for high spectral resolution NV-ESR.
The result of this experiment is shown in Fig.~\ref{fig:SingleNVData}(d). Clear reductions in intensity are observed compared to a sequence without the MW2 pulse.
The obtained spectrum agrees very well with the three resolved peaks and the spectral splitting of the P1 center.~\cite{Stepanov2016,Loubser1978} 
In addition, the $\sim$2 MHz linewidth of the observed peaks is of similar width to high resolution, inhomogeneously broadened, NV-ESR signals resolved at low magnetic fields.~\cite{Fortman2019}

%
%
In summary, we have demonstrated NV-ESR of P1 centers at a Larmor frequency of 115 GHz and the corresponding magnetic field of 4.2 Tesla. 
We have also shown that the application of chirp pulses improves excitation bandwidth and population inversion of the NV center spin-state at high magnetic fields.
The high magnetic field achieved in this measurement represents a step towards high-resolution NV-based spectroscopy. 
In the future, the presented HF ODMR system and technique can be further extended for the operation at a ESR Larmor frequency of 230 GHz (corresponding to 8.2 Tesla) for higher spectral resolution.
HF NV-ESR offers insight into complex radical spin systems, with high spectral resolution and the capability of the detection of nanoscale heterogeneity of external spins, spectral separation from unwanted ESR signals such as diamond surface spins, and applications for the study of spins within complex intracellular environments.
Furthermore, the present demonstration sets the basis for HF NV-detected NMR spectroscopy enabling high resolution NMR of a small number of molecules in spatially and temporally heterogeneous environments. 
HF NV-detected NMR spectroscopy will be useful for a variety of investigations including structures and dynamics of biomacromolecules and chemical environments of solid state surfaces and interfaces.

%
%
This work was supported by the National Science Foundation (DMR-1508661 and CHE-1611134), the USC Anton B. Burg Foundation and the Searle scholars program (ST). Steffen J. Glaser is acknowledged for helpful discussion regarding pulse shaping.

See the supplementary material for the X-band ESR characterization of the ensemble NV sample.

%
%
\smallskip
\noindent{\bf DATA STATEMENT}\newline
The data that support the findings of this study are available from the corresponding author upon reasonable request.

%

%
%

\end{document}